\documentclass[11pt]{article}
\usepackage{lineno}
\usepackage{multirow}
\usepackage{enumitem}
\usepackage{amsmath}
\usepackage{float}
\usepackage{graphicx} 
\usepackage{amssymb,amscd,amsmath,amsthm,color}
\usepackage[T1]{fontenc} 
\usepackage{subcaption}
\usepackage{enumitem}
\usepackage{mathtools}
\usepackage{tcolorbox}
\usepackage{verbatim}
\usepackage{mathrsfs}
\usepackage{physics}
\usepackage{mathtools}
\usepackage{slashed}
\usepackage{longtable}
\usepackage{jheppub}
\usepackage{hyperref}
\usepackage{cleveref}

\preprint{USTC-ICTS/PCFT-26-04}

\def\be{\begin{equation}}
\def\ee{\end{equation}}
\def\bea{\begin{eqnarray}}
\def\eea{\end{eqnarray}}
\def\dd{\mathrm d}

\numberwithin{equation}{section}

\makeatletter
\g@addto@macro\bfseries{\boldmath}\g@addto@macro\bfseries{\boldmath}
\makeatother
\title{Index saddle for supersymmetric F1--P  black ring}

\author{Pavan Dharanipragada$^a$,
    Gurmeet Singh Punia$^{b,c}$,    Amitabh Virmani$^d$}
\emailAdd{pavan@physics.iitm.ac.in, gurmeet96@ustc.edu.cn, avirmani@cmi.ac.in}
\affiliation[a]{Centre for Strings, Gravitation and Cosmology,
Department of Physics,
Indian Institute of Technology Madras, Chennai, India 600036}
\affiliation[b]{
Interdisciplinary Center for Theoretical Study,
University of Science and Technology of China, Hefei, Anhui 230026, China }
\affiliation[c]{
Peng Huanwu Center for Fundamental Theory, Hefei, Anhui 230026, China}
\affiliation[d]{Chennai Mathematical Institute, 
H1, SIPCOT IT Park, Siruseri, Kelambakkam 603103 India}

\abstract{We construct the index saddle for the supersymmetric F1--P black ring. Our construction proceeds by taking a supersymmetric limit of a non-supersymmetric doubly spinning F1--P black ring. We express the resulting saddle as a three-center Bena--Warner solution. The black ring saddle possesses a finite-area event horizon, yet the two-derivative index vanishes. The solution is singular on certain subspaces of the horizon, where higher-derivative corrections are expected to become important. We argue that, once such corrections are taken into account, the solution can yield a finite result. In particular, we present a scaling analysis showing that the index agrees with the microscopic result, up to an overall numerical constant that cannot be fixed by the scaling argument alone. This analysis applies only within a restricted region of parameter space, whose full significance is not yet fully understood.}
\makeatletter
\gdef\@fpheader{}
\makeatother
\begin{document}
\allowdisplaybreaks
\maketitle
\flushbottom

\section{Introduction}

Recently, there has been significant progress in understanding the supersymmetric index as a gravitational partition function and in identifying the gravitational saddles that contribute to the index \cite{Cabo-Bizet:2018ehj,Iliesiu:2021are}. This has led to a flurry of activity focused on constructing index saddles for supersymmetric black holes and comparing the resulting contributions with black hole entropy \cite{Cassani:2019mms, Bobev:2020pjk,Hristov:2022pmo,H:2023qko,Anupam:2023yns,Boruch:2023gfn,Hegde:2023jmp,Chowdhury:2024ngg,Chen:2024gmc,Cassani:2024kjn,Hegde:2024bmb,Adhikari:2024zif,Boruch:2025qdq,Bandyopadhyay:2025jbc,Boruch:2025biv,Cassani:2025iix,Boruch:2025sie}. 

Small black holes provide a sharp testing ground for these ideas. Small black holes are singular solutions of the two-derivative supergravity equations of motion, as they have vanishing horizon area. Microscopically, however, they are described by BPS states with non-zero degeneracies. A class of examples is provided by the F1–P system (also known as the Dabholkar–Harvey system) \cite{Dabholkar:1989jt, Dabholkar:1990yf, Dabholkar:1995nc}, in which a fundamental heterotic string winds a circle $w$ times and carries $n$ units of momentum along the same circle. This system is BPS if the right-moving sector is kept in its ground state. It is expected that this system is described by a small black hole in the gravitational description \cite{Callan:1995hn, Sen:1995in}. The index saddles for such small black holes were constructed in \cite{Chowdhury:2024ngg, Chen:2024gmc}. 

A natural question is what happens when a large angular momentum $J$ is added to the system \cite{Lunin:2002qf, Iizuka:2005uv, Dabholkar:2006za}. It was argued in these works that, in five dimensions, adding sufficiently large angular momentum to a small black hole causes it to transition into a small black ring. The small black ring is characterized  by the angular momentum 
$J$ and a dipole charge 
$Q$ in addition to the winding  and momentum charges $ (w,n)$. The dipole charge $Q$ counts the winding of the string along the ring direction. 


On the microscopic side, it is now well known that both the degeneracy and index of an elementary heterotic string with the above charges behave such that their logarithm goes as\footnote{See Section 2 of \cite{Dabholkar:2006za} for a concise review.}
\be
S_\mathrm{micro} = 4 \pi \sqrt{n w - J Q}. \label{micro}
\ee
Using a scaling argument similar to the one used for small black holes \cite{Sen:1995in}, it was shown in \cite{Dabholkar:2006za} that the entropy of the small black ring scales as, 
\be
S_\mathrm{macro} = C \sqrt{n w - J Q}, \label{macro}
\ee
for some constant $C$. While the numerical constant cannot be determined from the scaling analysis alone, it was argued that the same constant appears in the scaling of the entropy of the four-dimensional small black hole.

There is, however, a  caveat in this reasoning.  While at weak coupling, the index and the degeneracies are the same, at strong coupling, where the black ring description is applicable, there is no argument guaranteeing the equality of the degeneracy and the index. The scaling analysis applies strictly to the degeneracy, whereas the protected quantity is the index. To circumvent this issue, one must provide a scaling argument for the index. To do so, one must construct an appropriate index saddle. 


In this paper, we construct the index saddle for the small black ring carrying all four charges 
$n, w, J, Q$. Ref.~\cite{Bandyopadhyay:2025jbc} constructed the index saddle for a small black ring that did not carry an independent dipole charge. This limitation arose for purely technical reasons, which can be understood as follows.

In the approach taken by Anupam, Chowdhury, and Sen \cite{Anupam:2023yns}, and subsequently adopted by later authors (for example \cite{Chowdhury:2024ngg, Chen:2024gmc}), the construction of index saddles typically proceeds by taking a supersymmetric limit of a family of non-extremal solutions. For complicated systems such as black rings, such families of solutions are  generally not known, and when they can be constructed, the required techniques are exceedingly intricate. In contrast, another approach to constructing index saddles, pursued in \cite{Boruch:2025biv, Cassani:2025iix, Boruch:2025sie}, is based on multi-center supersymmetric solutions. This approach is supersymmetric from the outset and therefore does not require taking any limit.

The hurdle is for small black holes the applicability of the second approach  is not readily clear. One encounters conceptual puzzles that cannot be resolved in a simple way;  for example, although the new attractor mechanism \cite{Boruch:2023gfn} can  be used to fix the charges at the two centers of the index saddle \cite{Chen:2024gmc} for small holes, naively the results look surprising.\footnote{See comments on pages 25-26 of \cite{Chen:2024gmc}.} For small black rings similar issues are expected to arise. We therefore believe that the first approach is better suited to the problem: it is technically viable and appears to be conceptually better controlled. After all, for small black holes it gave very promising results \cite{Chowdhury:2024ngg, Chen:2024gmc}.

In following the first approach, one must therefore take a supersymmetric limit of a black ring carrying two angular momenta, two electric charges, and a dipole charge.
We have recently succeeded in constructing the required non-extremal solution \cite{our-paper}. In this paper, we take the supersymmetric limit and construct the index saddle for the supersymmetric F1--P small black ring.

The saddle solution allows us to address at least partially the caveat in the scaling analysis mentioned above.  The mechanism by which this works is closely analogous to that for small black holes \cite{Chowdhury:2024ngg}. Following the algorithm for computing the index using the corresponding index saddle, one finds that, in the two-derivative theory, the logarithm of the index vanishes just as the entropy of the small black ring vanishes at the two-derivative level. However, despite possessing an event horizon of finite area, the small black ring index saddle is not completely smooth: it is singular on two circles of the horizon where higher-derivative corrections are expected to become important. We analyze the geometry in the vicinity of these singularities and, using the symmetries of the theory together with a scaling argument closely paralleling those employed in \cite{Chowdhury:2024ngg, Dabholkar:2006za}, determine the corrected logarithm of the index up to an overall numerical factor. The resulting charge dependence of the index is found to be in agreement with the microscopic result \eqref{micro}.   However, our scaling analysis only works when the black ring saddles  are  sufficiently ``thin'', as will be explained in the main text.

The rest of the paper is organized as follows. In Section \ref{sec:review}, we review the supersymmetric F1--P black ring whose index saddle we are seeking. In Section \ref{sec:index-saddle}, we present the index saddle. In Section \ref{sec:properties}, we analyse some simple properties of the index saddle.  In Section \ref{sec:BW}, we write the index saddle solution as a three-center Bena--Warner solution. 
In Section \ref{sec:scaling}, we present the scaling analysis. 
We close with a brief discussion in Section \ref{sec:conclusions}. Some technical details are relegated to appendix \ref{sec:appendix}, where a discussion on the relation between our present analysis to ref.~\cite{Bandyopadhyay:2025jbc} can also be found.

Throughout this work, we follow the conventions and notation of our previous paper \cite{our-paper}.

\section{Supersymmetric F1--P black ring}

\label{sec:review}

In this section, we review the supersymmetric F1--P black ring \cite{Elvang:2004ds, Elvang:2004xi, Dabholkar:2006za}. We follow the notation and conventions of \cite{our-paper}. The metric takes the form,
\bea
	\dd s_5^2&=&-(h_1 h_2)^{-2/3}(\dd t+k)^2+(h_1 h_2)^{1/3}\dd s_4^2, 
	\label{bpsmetric} \\
	h_i &=&1+ \frac{Q_i}{4 \varkappa^2}(x-y),\quad 
	k= - \frac{1}{2}q(1+y) \dd \psi, \label{h-omega}
\eea
with the dipole charge $q$ parameterized as
\begin{equation} \label{dipole-alpha}
    q=\frac{1}{2\sqrt{2} \varkappa} \sqrt{Q_1 Q_2} (\sqrt{1+\alpha}+ \sqrt{1-\alpha}),
\end{equation}
with $0 \le \alpha \le 1$. The sign of $q$  is positive in our conventions. 
The four-dimensional base metric $\dd s_4^2$ is flat space written in ring coordinates,
\begin{equation}
	\dd s_4^2= \frac{2\varkappa^2}{(x-y)^2} \left[\frac{\dd y^2}{y^2-1} + (y^2-1)\dd\psi^2 + \frac{\dd x^2}{1-x^2} + (1-x^2) \dd\phi^2 \right].
	\label{ds4}
\end{equation}
The $S^1$ of the ring lies lies along the $\psi$-direction. The angular coordinates $\phi$ and $\psi$ both have $2 \pi$ periodicity. The radial coordinate $y$  range over $- \infty < y \le -1$, and the polar coordinate $x$ on the $S^2$ ranges over $- 1 \le x \le 1$. The horizon of the black ring is located at $y \to -\infty$. 

The remaining fields supporting the solution are,
\begin{align}
&	B =- \frac{1}{2 h_2 } q(1+y) \dd t \wedge \dd \psi,& &
e^{2\Phi} =\frac1{h_2},&    & 
    e^{- \sqrt{\frac{3}{2}}\chi} =  \frac{h_1}{\sqrt{h_2}}, \\
&	A^{(1)} =\dd t - h_1^{-1}(\dd t + k),  & 
&	A^{(2)}=\dd t  - h_2^{-1}(\dd t + k),  & &
    A^{(3)} = \frac{1}{2} q (1 - x) \dd \phi, 
\end{align}
where
\be \label{A3-def}
e^{-\frac{\sqrt{2}}{\sqrt{3}}\chi - 2\Phi} \star_5 H =: - d A^{(3)} \qquad  \mbox{and}  \qquad H = \dd B - \dd A^{(2)} \wedge A^{(1)}. 
\ee
The angular momentum of the black ring is related to its dipole charge via,
\begin{equation} \label{J-psi-small-black-ring}
J := J_\psi=\frac{\pi }{4 G_5} (2 \varkappa^2) q.
\end{equation}

It is useful to express the supersymmetric solutions in the Bena–Warner formalism \cite{Bena:2007kg}.
 The eight Bena--Warner  functions characterizing the  black ring are, 
\begin{align} \label{BW-functions-F1--P-small}
L_1 &= 1 + \frac{Q_1}{4 r_o}, & K^1 &=0,  &
L_2 &= 1 +  \frac{Q_2}{4 r_o},&  K^2 &=0, \\
L_3 &= 1 & K^3 &= \frac{q}{2 r_o}, &
V &= \frac{1}{r}, & M &=- \frac{q}{4}+  \frac{q \varkappa^2}{8r_o},
\end{align}
where
\be
r_o := | \vec x - \vec x_o|,  \qquad  \mbox{with} \qquad  \vec x_o = \left(0,0, -\frac{1}{2} \varkappa^2\right).
\ee
On the three-dimensional base space, the ring is located at $\vec x =\vec x_o$.

We interpret this configuration as a solution of heterotic supergravity  compactified on $T^4 \times S^1$. Let the $S^1$ be  labeled by $z$ and have circumference $2 \pi R_z$. The $T^4$  will not play any role in our discussion.  The parameters $Q_1, Q_2, q$ and $\varkappa^2$ are related to the the quantized charges $n,w,Q$ and angular momentum $J$ via the relations \cite{Dabholkar:2006za} (with $\alpha' = 1$),
\begin{align} \label{quantized}
 q& = \frac{g^2 Q}{R_z}, & \varkappa^2 &= \frac{J}{2Q}, & Q_1 &= \frac{g^2 n}{R_z^2}, & Q_2 &= g^2 w.    
\end{align}
Here $n$ and $w$ denote the momentum and winding numbers along the  $S^1$. The parameter $Q$  represents the winding number along the $S^1$ of the ring. The fact that $\alpha$ in \eqref{dipole-alpha} ranges over $0\le \alpha \le 1$ puts  an upper bound on the angular momentum. This bound translates into 
\be
n w - JQ \ge 0. 
\ee
This black ring is a small black ring, i.e., as a solution to the two-derivative supergravity theory its horizon area and hence the Bekenstein-Hawking entropy vanish. The physical electric charges as defined in \cite{our-paper} are related to the 
charge parameters $Q_{1,2}$ as,
\be
\mathbf{Q}_{1,2} = \frac{\pi}{4 G_{5}} Q_{1,2}. 
\ee
The black ring saturates the BPS bound $
M = \mathbf{Q}_{1} + \mathbf{Q}_{2}.$

\section{Index saddle for the supersymmetric F1--P black ring}
\label{sec:index-saddle}

In this section, we construct the dominant gravitational index saddle for the supersymmetric F1--P black ring. As mentioned in the introduction, our construction proceeds by taking a carefully chosen supersymmetric limit of a non-extremal solution. The required non-extremal charged F1--P black ring solution was recently constructed in \cite{our-paper}. The full solution has six parameters: \be
a, b, c, \delta_1, \delta_2, \varkappa.
\ee
The parameter $\varkappa$ sets the overall scale of the solution; all  other parameters are dimensionless.  
The boost parameters $\delta_1$ and $\delta_2$ are related to the electric P and F1  charges,  respectively. The parameter $c$ determines the location of the horizon, which lies at $y=-1/c$. The parameter $a$ is related to the dipole charge\footnote{More precisely, the difference $a-c$ is related to the seed dipole charge.}, while the parameter $b$ controls the rotation of the ring along the $S^2$ cross-section.

This intuitive understanding of the parameters is reliable only when they take small values.
For example, for 
$b \ll 1$ one finds that 
$J_\phi$  is proportional to $\sqrt{b}$. When 
$b$ is of order unity, however, the rotation on the 
$S^2$ is no longer related to 
$b$ in any simple way. More generally, it is difficult to develop an intuitive understanding of how these parameters map to physical quantities. For this reason, it is not straightforward to take a limit in the full parameter space that yields the index saddle solution.

After an extensive search, we find that the correct limit to consider is 
\be
b \to -1,
\ee
keeping $1+b$ positive throughout the limit\footnote{Some intuitive explanation on why $b\to -1$ is the appropriate  limit can be found in appendix \ref{sec:appendix}.}, together with 
\be
\delta_1, \delta_2 \to \infty,
\ee
while keeping the electric charges finite. There are still some choices to be made. We take,
\bea
\delta_1 &=& \frac{1}{2} \sinh^{-1} \left[ \frac{Q_1}{(2a)(1+b) \varkappa^2} \right], \label{delta-1}\\ 
\delta_2 &=& \frac{1}{2} \sinh^{-1} \left[ \frac{Q_2}{(2a)(1+b) \varkappa^2} \right], \label{delta-2}
\eea
as $b\to -1$. The inclusion of  the factor $(2a)$ in the denominator of  \eqref{delta-1}--\eqref{delta-2} is optional, but it has the advantage that in the BPS limit the physical charges $\mathbf{Q}_{1,2}$,
as defined in \cite{our-paper}, are related to the charge parameters $Q_{1,2}$ in a particularly  simple way,
\be
\mathbf{Q}_{1,2} = \frac{\pi}{4 G_{5}} Q_{1,2}. 
\ee
The resulting configuration  saturates the BPS bound 
\be
M = \mathbf{Q}_{1} + \mathbf{Q}_{2}. \label{BPS-bound}
\ee

In the non-extremal Lorentzian solution \cite{our-paper}, the parameter $b$ is a positive number in the range,
\be
0 \le b < \frac{1-a}{1+a},
\ee
together with 
\be \label{bounds}
0 \le c \le a < 1 \qquad \mbox{and} \qquad \varkappa > 0.
\ee
Therefore, $b \to -1$ also corresponds to an analytic continuation of the solution. In particular, it introduces factors of the imaginary unit $\mathrm{i}$. For the index saddle, parameters $a, c$ and $\varkappa$ continue to satisfy \eqref{bounds}.

In the BPS limit, the five-dimensional Einstein frame metric simplifies to 
\bea
  \mathrm{d}s^2_5 &= & - (h_1 h_2)^{-\frac{2}{3}} (\mathrm{d} t +k_\psi \mathrm{d} \psi 
   + k_\phi \mathrm{d} \phi)^2 + (h_1 h_2)^{\frac{1}{3}}   ds^2_\mathrm{base},
\eea
with
\bea    \label{4d-base}
  ds^2_\mathrm{base} &=&  \frac{2 \varkappa^2 H  (x,y)  }{(x-y)^2} \left\{ \frac{F  (x,y) \mathrm{d}\psi^2}{H  (x,y)^2} 
  - \frac{G(x) G(y) \, \mathrm{d}\phi^2}{F  (x,y)}+ \frac{1}{4} \left[ \frac{\mathrm{d}x^2}{G(x)} - \frac{\mathrm{d}y^2}{G(y)} \right] \right\}.
\eea
and
\bea
k_\psi &=& \frac{\sqrt{Q_1 Q_2} (1+y)}{4 \varkappa \sqrt{2a}  H  (x,y)} \left( \sqrt{a+c} \, J_+ (x,y) + \sqrt{a-c} \, J_-(x,y) \right), \\
k_\phi &=& - \mathrm{i}  \sqrt{Q_1 Q_2}  (1-x^2) \frac{c (c + 2 y + cy^2)}{4 \varkappa \sqrt{2 a} H  (x,y)}    \sqrt{1-a^2} \left( \sqrt{a+c} - \sqrt{a-c} \right).
\eea
Note the appearance of the imaginary unit in $k_\phi$, as expected from the analytic continuation mentioned above. The functions $G$, $H$, $F$,  and $J_\pm$ are obtained by taking the $b \to -1$ limit of the corresponding functions of Chen, Hong, and Teo \cite{Chen:2012kd}. They take the form
\begin{align}
& G(x) = (1-x^2)(1+c x), \\[2mm]
& H(x,y) = (2 + c(x+y))^2 - c^2 (1-x y)^2,\\[2mm]
& F  (x,y) = \left(y^2-1\right) (1+ c x) \left(c^2 \left(1-x^2\right)
   \left(y^2-1\right)+4 (1+ c x) (1+c y)\right), \\[2mm]
& J_\pm(x,y) = \pm a c (c + 2 x + c x^2) (1-y) + 
 c^2 (1 - x) (1 - x y) - (2 + c + c x) (2 + c (x + y)), \nonumber \\
\end{align}
Note that in the $b\to -1$ limit the function $H(x,y)$ is such that $H(x,y) = H(y,x)$ and is also equals to the function $K(x,y)$ of Chen, Hong, and Teo. The  functions $h_{1,2}$ are
\begin{align} \label{functions-h1}
	h_1 &= 1 + \frac{Q_1 (x-y) \{ 2 + c (x + y) - c^2 (1 - x y) \}}{2 \varkappa^2 H  (x,y)}, \\
    \label{functions-h2}
	h_2 &= 1 + \frac{Q_2 (x-y) \{ 2 + c (x + y) - c^2 (1 - x y) \}}{2 \varkappa^2 H  (x,y)}.
\end{align}

The remaining fields supporting the solution are as follows. The scalars are
\begin{align} \label{scalars-final}
	e^{2\Phi} &= \frac{1}{h_2}, &  e^{-\tfrac{\sqrt{3}}{\sqrt{2}} \chi} &= \frac{h_1}{\sqrt{h_2}}.
\end{align}
The two vectors are,
\begin{align}
	A^{(1)}_{t} & = 1- h_1^{-1} , &
	A^{(1)}_{\phi} & = - h_1^{-1} k_\phi, &
	A^{(1)}_{\psi} & =  - h_1^{-1} k_\psi,\\
	A^{(2)}_{t} & = 1- h_2^{-1} , &
	A^{(2)}_{\phi} & = - h_2^{-1} k_\phi,&
	A^{(2)}_{\psi} & = - h_2^{-1} k_\psi,
\end{align}
and finally the B-field components are,
\begin{align}
	B_{t \phi} & = h_2^{-1} k_\phi, &
	B_{t \psi} & = h_2^{-1} k_\psi. 
\end{align}
The solution above provides the gravitational index saddle for the F1--P black ring reviewed in Section~\ref{sec:review}, as we now confirm.

 We can readily compute the physical quantities of interest from the metric given above. The inverse temperature $\beta = T^{-1}$ and the angular velocity $\Omega_\phi$ of the solution are
\begin{align} \label{beta}
	\beta = T^{-1} & = \frac{\pi}{2 \varkappa} \sqrt{\frac{{Q}_1 {Q}_2 (1-a^2)}{2 a} } \left( \sqrt{a+c} - \sqrt{a-c}\right),
\end{align}
and
\begin{align}
	\Omega_\phi & = 4 \mathrm{i} \varkappa \sqrt{\frac{2 a}{(1-a^2) {Q}_1 {Q}_2}} \frac{1}{\left( \sqrt{a+c} - \sqrt{a-c} \right)}.
\end{align}
These quantities satisfy
\be
\beta \Omega_\phi = 2\pi \mathrm{i}. \label{Omega-beta}
\ee
This relation confirms that the chemical potential $\Omega_\phi$ for the angular momentum $J_\phi$ is adjusted so as to insert a factor
\be
e^{-\beta \Omega_\phi J_\phi} = e^{-2\pi \mathrm{i} J_\phi} = (-1)^F
\ee
in the gravitational path integral. Therefore, the doubly rotating black ring provides the dominant saddle contributing to the index.

The area of the horizon $A$ is
\begin{align}
\label{A-H}
	A & = \pi^2 \, c\,\varkappa \sqrt{\frac{2 {Q}_1 {Q}_2 (1-a^2) }{a}} \left( \sqrt{a+c} - \sqrt{a-c} \right),
\end{align}
and the $S^2$ angular momenta $J_{\phi}$ is
\begin{align} \label{J-phi}
	J_\phi & = \mathrm{i} \frac{\pi c\,\varkappa}{4 G_5} \sqrt{\frac{{Q}_1 {Q}_2 (1-a^2)}{2a}} \left( \sqrt{a+c} - \sqrt{a-c} \right).
\end{align}
These quantities satisfy 
\be
S+2 \pi \mathrm{i} J_{\phi} = 0. \label{S-J}
\ee
This relation confirms that  the gravitational index computed using the saddle solution vanishes.

At this point, it is useful to recall that for any two-derivative theory in five dimensions containing the metric, a two-form field, vector fields, and scalars, there exists a scaling symmetry
\begin{align}
& g_{\mu \nu} \to \lambda^2 g_{\mu \nu}, & & B_{\mu \nu} \to \lambda B_{\mu \nu}, & & A_\mu \to \lambda A_\mu, & & \phi \to \phi, \label{scaling}
\end{align}
under which the action scales as $\lambda^3$. For a classical black hole solution of such a theory, the mass $M$, angular momenta $J_\phi, J_\psi$, dipole charge $q$, and electric charges $Q_i$, scale under \eqref{scaling} as
\begin{align}
& M \to \lambda^2 M, & & J_{\phi,\psi} \to \lambda^3 J_{\phi,\psi}, & & Q_i \to \lambda^2 Q_i, & & q \to \lambda q. \label{scaling-charges}
\end{align}
Under this scaling, the entropy of the black hole scales as
\be
S \to \lambda^3 S. \label{scaling-entropy}
\ee

In the solution written above, the parameters $a$ and $c$ are dimensionless, while the parameter $\varkappa$ sets the overall scale. Under the scaling $\varkappa \to \lambda \varkappa$, we see from \eqref{delta-1}–\eqref{delta-2} that the electric charges $Q_1$ and $Q_2$ scale as in \eqref{scaling-charges}. Similarly, one can verify that the scaling of the entropy in \eqref{A-H} and of the angular momentum in \eqref{J-phi} is consistent with \eqref{scaling-charges}--\eqref{scaling-entropy}.

Note, however, that the same scaling leads to only a factor of $\lambda^2$ when applied to the microscopic entropy \eqref{micro}, which appears to be in contradiction with the above result. This apparent discrepancy is resolved by the fact that, in two-derivative supergravity, the classical black hole has vanishing entropy, as noted at the end of the previous section. Equation \eqref{S-J} confirms that not only the entropy but also the gravitational index computed using the saddle solution vanishes in the two-derivative theory.

The $\Omega_{\psi}$ angular velocity is zero for the solution. The angular momenta $J_{\psi}$ is,
\begin{align} \label{J-psi-saddle}
	J_\psi & = \frac{\pi \varkappa}{4 G_5} \frac{\sqrt{{Q}_1 {Q}_2}}{\sqrt{2a}} \left( \sqrt{a+c} \, (1+ a c) + \sqrt{a-c} \,  (1- a c) \right),
\end{align}
and the dipole charge $q$ is,
\begin{align} \label{dipole-saddle}
	q & = \frac{1}{2\varkappa}\frac{\sqrt{{Q}_1 {Q}_2}}{\sqrt{2a}} \left( \sqrt{a+c} + \sqrt{a-c}\right) .
\end{align}

The index saddle has one additional parameter compared to the small supersymmetric F1--P black ring. From a physical perspective, it is natural to interpret this extra parameter as $\beta$, the size of the thermal circle at infinity. For practical calculations, however, it is most convenient to take $a$ as the additional parameter and write $c = a \alpha$. From~\eqref{bounds} it follows that $0 \le \alpha \le 1$. Taking the limit $a \to 0$, we recover the supersymmetric F1--P black ring solution reviewed in Section~\ref{sec:review}.  
This property, together with equations~\eqref{BPS-bound}, \eqref{Omega-beta}, and \eqref{S-J}, confirms that the solution presented in this section is the index saddle for the F1--P black ring.   However, in the limit $a \to 0$, the inverse temperature \eqref{beta} does not diverge. This remains an ill-understood aspect of small black hole index saddles \cite{Chen:2024gmc, Bandyopadhyay:2025jbc}.



\section{Properties of the saddle solution}
\label{sec:properties}

In this section, we study some basic properties of the index saddle constructed in the previous section. Our first observation is that the four-dimensional base metric \eqref{4d-base} is simply flat space. It can be written as
\be \label{r1r2-base}
ds^2 = \dd r_1^2 + \dd r_2^2 + r_1^2 \dd \phi^2 + r_2^2 \dd \psi^2,
\ee
where
\bea \label{r1-change}
r_1^2 &=& \frac{2 \varkappa^2 (1-x^2)(1+ c y)}{(x-y)^2}, \\
r_2^2 &=& \frac{2 \varkappa^2 (y^2 -1)(1+ c x)}{(x-y)^2}.
 \label{r2-change}
\eea
Note that the parameter $a$ does not appear in the base metric. When $c=0$, the coordinates $(x,y)$ are such that the base metric reduces to the form given in \eqref{ds4}.

A direct calculation shows that the functions $h_{1,2}$ defined in \eqref{functions-h1}--\eqref{functions-h2} are harmonic on the four-dimensional base space. The horizon of the black ring is located at $y=-1/c$. Motivated by the discussion in \cite{Bandyopadhyay:2025jbc}, it is natural to expect that the sources for the harmonic functions $h_1$ and $h_2$ are supported on the $S^2$ cross-section of the horizon, localized at its north and south poles.

When described in terms of the base-space coordinates, these source locations correspond to two concentric circles: an inner circle associated with the north poles of the $S^2$ cross-section, and an outer circle associated with the south poles. In \cite{Bandyopadhyay:2025jbc}, these two circles were denoted using a $\pm$ notation. Here, we adopt the  terminology of north and south circles and define
\bea
\text{North circle}: && x = +1, \qquad y = -1/c,\\
\text{South circle}: && x = -1, \qquad y = -1/c.
\eea
From  \eqref{r1-change}--\eqref{r2-change}, we have
\bea
\text{North circle}:  \quad r_1 = 0, \quad r_2 = \sqrt{2}\,\varkappa \sqrt{1-c}, \\
\text{South circle}:  \quad r_1 = 0, \quad r_2 = \sqrt{2}\,\varkappa \sqrt{1+c}.
\eea

For a uniform ring source located at $r_1 = 0$, $r_2 = R$ in the $(r_1,\phi,r_2,\psi)$ coordinates, the relevant harmonic function is $\Sigma^{-1}$ \cite{Emparan:2006mm}, where
\be
\Sigma = \sqrt{(r_1^2 + r_2^2 + R^2)^2 - 4 R^2 r_2^2}.
\ee
For uniform ring sources smeared on the north and the south circles defined above the relevant  harmonic functions $\Sigma_N^{-1}$ and $\Sigma_S^{-1}$ are  therefore, 
\bea
\frac{1}{\Sigma_N} &=& \frac{1}{2 \varkappa^2}  \left[\frac{(x-y)}{2 - c (1 - x - y - x y)}\right]  \implies \Sigma_N = \frac{2 \varkappa^2}{x-y} \{2 - c (1 - x - y - x y)\},\\
\frac{1}{\Sigma_S} &=& \frac{1}{2 \varkappa^2}   \left[\frac{(x-y)}{2 + c (1 + x + y - x y)}\right] \implies \Sigma_S = \frac{2 \varkappa^2}{x-y} \{2 + c (1 + x + y - x y)\}.
\eea
Note that in the $c=0$ limit, both $\Sigma_N$ and $\Sigma_S$ reduce to
\be
\Sigma = \frac{4\varkappa^2}{x-y},
\ee
which appears in the harmonic functions $h_{1,2}$ of the small F1--P black ring, cf.~\eqref{h-omega}.

In terms of $\Sigma_{N,S}^{-1}$ the harmonic functions $h_{1,2}$ for the index saddle are 
\bea
h_1 &=& 1 + \frac{1}{2}\frac{Q_1(1-c)}{\Sigma_N} + \frac{1}{2}\frac{Q_1(1+c)}{\Sigma_S}, \\
h_2 &=& 1 + \frac{1}{2}\frac{Q_2(1-c)}{\Sigma_N} + \frac{1}{2}\frac{Q_2(1+c)}{\Sigma_S},
\eea
confirming that the sources for these harmonic functions are located at the north and south circles. Moreover, these expressions realize the ``splitting-centers'' picture proposed in \cite{Boruch:2025qdq}: the charges $Q_i$ are split between the two circles as $\frac{1}{2}(1-c)Q_i$ and $\frac{1}{2}(1+c)Q_i$. These observations naturally motivate a reformulation of the index saddle in the Bena--Warner framework, which we carry out in the next section.

\section{Index saddle in the Bena--Warner form} 
\label{sec:BW}

To set the notation, we recall that in the Bena--Warner formalism~\cite{Bena:2007kg} the five-dimensional Einstein-frame metric takes the form
\be
\dd s^2 = -f^2(\dd t + k )^2 + f^{-1} \dd s^2_{\mathrm{4d\text{-}base}}, \label{5d-metric}
\ee
with the four-dimensional base metric $\dd s^2_{\mathrm{4d\text{-}base}}$ written in the Gibbons--Hawking form as
\begin{equation}
	\dd s^2_{\mathrm{4d\text{-}base}} = V^{-1}(\dd \widetilde z + A)^2 + V\,\dd s^2_{\mathrm{3d\text{-}base}}.
	\label{GibbHawk}
\end{equation}
In equation \eqref{GibbHawk}, $\dd s^2_{\mathrm{3d\text{-}base}}$ is the flat three-dimensional base metric and $\widetilde z$ denotes the coordinate along the Gibbons--Hawking fiber. $V$ is a harmonic function on the three-dimensional flat base space. The Bena--Warner solutions are specified in terms of eight harmonic functions $\{V, K^I, L_I, M\}$, with $I = 1,2,3$.

The three-dimensional one-form $A$ satisfies $\star_3 \dd A = \dd V $, 
where $\star_3$ denotes the Hodge star on the three-dimensional base space. The one-form $k$ on the four-dimensional base space is given as
\be
k = \mu(\dd \widetilde z + A) + \omega_3, \label{k-1} \qquad 
\mu = \frac{1}{6} C_{IJK} \frac{K^I K^J K^K}{V^2} + \frac{1}{2V} K^I L_I + M,
\ee
with $C_{IJK} = 1$ if $(IJK)$ is a permutation of $(123)$ and $C_{IJK} = 0$ otherwise. The three-dimensional one-form $\omega_3$ satisfies
\be
\star_3 \dd \omega_3 = V \dd M - M \dd V + \frac{1}{2}\left(K^I \dd L_I - L_I \dd K^I\right).
\label{omega-2}
\ee
The function $f$ appearing in \eqref{5d-metric} takes the form $f = (h_1 h_2 h_3)^{-1/3}$, where the three functions $h_I$ are defined as
\be
h_I = \frac{1}{2V} C_{IJK} K^J K^K + L_I.
\ee
Further details can be found in \cite{Adhikari:2024zif,our-paper}.

With the coordinate transformations,
$r_1 = \rho \cos \Theta,  r_2 = \rho \sin \Theta,   \phi = \frac{1}{2} \left(\phi_1  + \phi_2\right),$  $\psi = \frac{1}{2} \left(\phi_1  - \phi_2\right),
$
followed by $\Theta = \frac{1}{2} \theta, 
\rho = 2 \sqrt{r},$ the four-dimensional flat base space \eqref{r1r2-base} can be written in the Gibbons--Hawking form as
\be
\dd s^2_4 = r ( \dd \phi_1 + \cos \theta \dd \phi_2)^2 + \frac{1}{r} ( \dd r^2  + r^2 \dd \theta^2 + r^2 \sin^2 \theta \dd \phi_2^2). \label{4d-flat-GH}
\ee In this form it is clear that the Bena--Warner harmonic function $V$ is simply $1/r$. It is convenient to work with cartesian coordinates on the three-dimensional flat base space
$
x_1 = r \sin \theta \cos \phi_2, 
x_2 = r \sin \theta \sin \phi_2, 
x_3 = r \cos \theta.
$
In these coordinates, 
\begin{align}
  & r_N := | \vec x - \vec x_N| = \frac{1}{4}\Sigma_N,  & &  \mbox{where} & &  \vec x_N = \left(0,0, -\frac{1}{2} (1-c) \varkappa^2\right), & \\
 & r_S := | \vec x - \vec x_S| = \frac{1}{4}\Sigma_S,  & &\mbox{where} & &    \vec x_S = \left(0,0, -\frac{1}{2} (1+c) \varkappa^2\right). &
\end{align}

Now, one can verify that the following harmonic functions give the saddle solution in the form presented above,
\begin{align}
&L_1 = 1 + \frac{(1-c)Q_1}{8r_N}+ \frac{(1+c)Q_1}{8 r_S},&
&L_2 = 1 + \frac{(1-c)Q_2}{8r_N}+ \frac{(1+c)Q_2}{8 r_S},& \\
&L_3 = 1, &
&V = \frac{1}{r} , \\
&K^1 = 0, & 
&K^2 = 0, & \\
&K^3 = \frac{k^3_N}{r_N} + \frac{k^3_S}{r_S}, & 
&M = m_0 + \frac{m_N}{r_N} + \frac{m_S}{r_S}.  
\end{align}
with coefficients
\bea
&& k^3_N = \frac{\sqrt{Q_1 Q_2}}{8 \varkappa \sqrt{2a}}  
\left\{\left(1+a - \mathrm{i} \sqrt{1-a^2}\right)\sqrt{a-c}+ \left(1-a + \mathrm{i} \sqrt{1-a^2}\right)\sqrt{a+c} \right\},\\
&&k^3_S = \frac{\sqrt{Q_1 Q_2}}{8 \varkappa \sqrt{2a}}  
\left\{\left(1-a +\mathrm{i} \sqrt{1-a^2}\right)\sqrt{a-c}+ \left(1+a - \mathrm{i} \sqrt{1-a^2}\right)\sqrt{a+c} \right\},\\
&&m_0 = - \frac{\sqrt{Q_1 Q_2}}{8 \varkappa \sqrt{2 a}} \left\{ \sqrt{a+ c} + \sqrt{a-c}\right\},\\
&&m_N = \frac{(1-c)\sqrt{Q_1 Q_2}\varkappa}{32 \sqrt{2a}}  
\left\{\left(1+a + \mathrm{i} \sqrt{1-a^2}\right)\sqrt{a-c}+ \left(1-a - \mathrm{i} \sqrt{1-a^2}\right)\sqrt{a+c} \right\}, \nonumber \\ \\ 
&&m_S = \frac{(1+c)\sqrt{Q_1 Q_2}\varkappa}{32 \sqrt{2a}}  
\left\{\left(1-a -\mathrm{i} \sqrt{1-a^2}\right)\sqrt{a-c}+ \left(1+a + \mathrm{i} \sqrt{1-a^2}\right)\sqrt{a+c} \right\}. \nonumber \\
\eea
Upon setting $c = a \alpha$ and taking the limit $a \to 0$, these harmonic functions reduce to the harmonic functions \eqref{BW-functions-F1--P-small}. When the parameter $a$ is set equal to $c$, the same harmonic functions match those of the index saddle constructed in \cite{Bandyopadhyay:2025jbc}; see appendix~\ref{sec:appendix} for details.

Note that the sum $k^3_N + k^3_S$ is real. In fact, $k^3_N + k^3_S$ is proportional to the dipole charge \eqref{dipole-saddle}. In contrast, the sum $m_N + m_S$ is not real. In the Bena--Warner description, the harmonic function $M$ captures the momentum along the $\phi_1 = \psi + \phi$ direction. Indeed, it is straightforward to verify that $m_N + m_S$ is proportional to $J_\psi + J_\phi$. Since $J_\phi$ is purely imaginary for our index saddle, it is therefore expected that the total momentum charge captured by $m_N + m_S$ is not real. This feature represents one of the key differences between our approach and that of \cite{Boruch:2025sie}, which is closely tied to a four-dimensional description \cite{Boruch:2025qdq, Boruch:2025biv}, in which all total charges captured by the harmonic functions are taken to be real.

\section{Scaling analysis}
\label{sec:scaling}

The small black ring index saddle constructed in Section \ref{sec:index-saddle} is not completely smooth. In this section, we analyze the geometry in the vicinity of the resulting singularities. For this purpose, it is most convenient to work in the six-dimensional string frame. As follows from \eqref{scalars-final}, the dilaton vanishes on both the north and south circles, rendering the string-frame metric singular on these subspaces of the horizon. In these regions, higher-derivative corrections are therefore expected to become important.

In what follows, we zoom in near the north and south circles. Using the symmetries of the theory together with a scaling argument closely paralleling those employed in \cite{Chowdhury:2024ngg, Dabholkar:2006za}, we analyze the near-singularity geometry. This allows us to determine the corrections to the gravitational index. The resulting charge dependence is found to agree with the microscopic result \eqref{micro}. This agreement holds only within a restricted region of parameter space.

The region of the parameter space 
 where the match with the microscopic result is expected can be appreciated as follows.  Using $ c = a \alpha$, we note that the dipole charge for the index saddle \eqref{dipole-saddle} becomes 
\begin{equation} \label{dipole-alpha-2}
    q=\frac{1}{2\sqrt{2} \varkappa} \sqrt{Q_1 Q_2} (\sqrt{1+\alpha}+ \sqrt{1-\alpha}),
\end{equation}
which is the same as \eqref{dipole-alpha}---the dipole charge of the small black ring. However, the angular momentum $J_\psi$ of the index saddle 
\eqref{J-psi-saddle},
\be
	J_\psi = \frac{\pi \varkappa}{4\sqrt{2} G_5} \sqrt{Q_1 Q_2} \bigg\{ (\sqrt{1+\alpha} + \sqrt{1-\alpha})   + \alpha( \sqrt{1+\alpha}   -  \sqrt{1-\alpha})  a^2) \bigg\},
\ee
is not the same as the $J_\psi$ of the small black ring, which satisfies \eqref{J-psi-small-black-ring}. We do not understand the full significance of this observation. Though, we note that in an  expansion near $a =0$, the two expressions are the same to order $\mathcal{O}(a)$ but differ at $\mathcal{O}(a^2)$. Thus, only to $\mathcal{O}(a)$ we expect a match between the index computed using a scaling analysis with the microscopic answer.\footnote{As noted at the end of Section~\ref{sec:index-saddle}, small $a$ does not correspond to a diverging inverse temperature.
}  This is what we will show in  Section \ref{sec:limitations}. In Section \ref{near-singularities}, we analyse the geometry near the singularities for arbitrary values of $a,c$. 


\subsection{Geometry near the singularities}
\label{near-singularities}
The six-dimensional string-frame metric can be written as\footnote{These uplifts are discussed in \cite[Section 2]{our-paper}.}
\be 
\label{6d-string-final}
\dd s^2_{6S} = \frac{1}{h_2} \left\{-( \dd t + k)^2  + (\dd z -k)^2 + (h_1 -1) (\dd t + \dd z)^2 \right\} + \dd s^2_4.
\ee
In our presentation so far, the asymptotic value of the dilaton $\Phi$ has been set to zero. For the scaling analysis, the string coupling $g$ must be restored, so that the asymptotic value of the dilaton  $e^{\Phi}$ becomes $g$. Since shifting $\Phi$ by a constant is a symmetry of the string-frame equations of motion, this can be achieved simply by multiplying $e^{\Phi}$ by an overall factor of $g$, 
\be
\exp\left[2\Phi\right] = g^2 h_2^{-1}.
\ee
The B-field in six dimensions is 
\be
B = h_2^{-1} \, (\dd t + \dd z) \wedge k+ (1 - h_2^{-1}) \, \dd t \wedge \dd z .
\ee

\subsubsection{North circle}
We obtain the geometry in the vicinity of the north circle via the Bena--Warner formalism, following \cite{our-paper}.  To zoom in, we  take 
\be
r_N \ll Q_1, Q_2, \varkappa^2. \label{inequalities}
\ee
In this limit, the harmonic functions become,
\begin{align}\label{limit-N-1}
&L_1 \simeq  \frac{(1-c)Q_1}{8r_N},&
&L_2 \simeq \frac{(1-c)Q_2}{8r_N}& 
&L_3 = 1, &
&V \simeq \frac{2}{(1-c)\varkappa^2} , \\
&K^1 = 0, & 
&K^2 = 0, & 
&K^3 \simeq \frac{k^3_N}{r_N}, & 
&M \simeq \frac{m_N}{r_N}.  \label{limit-N-2}
\end{align}
Since the Bena--Warner function $V$ is now a constant, the four-dimensional base space becomes $\mathbb{R}^3 \times S^1$. The base metric is
\be \label{base-north}
\dd s^2_4 \simeq 2(1-c) \varkappa^2 \dd \psi_N^2 + \frac{2}{(1-c)\varkappa^2} \left(d r_N^2 + r_N^2 \dd \theta_N^2 + r_N^2  \sin^2 \theta_N \dd \phi_N^2\right),
\ee
where $(r_N,\theta_N,\phi_N)$ are spherical polar coordinates centered at $\vec x = \vec x_N$. In writing \eqref{base-north}, we have rescaled the fiber coordinate by a factor of 1/2 and called it $\psi_N$. $\psi_N$ has periodicity $2\pi$. Next, we define 
\be
\rho_N = \frac{\sqrt{2}}{\varkappa \sqrt{1-c}} r_N,
\ee
so that the base metric is
\be \label{base-simp}
\dd s^2_4 \simeq 2(1-c)\varkappa^2 \dd \psi_N^2 + \dd \rho_N^2 + \rho_N^2 ( \dd \theta_N^2 + \sin^2 \theta_N \dd \phi_N^2).
\ee

We can readily compute the one-form $k$ near the north circle, via \eqref{k-1}--\eqref{omega-2},
\be
k = \left(\frac{K^3}{2V} + M\right) (2 \dd \psi_N) + \omega_3. 
\ee
To find $ \omega_3$ we need to use the duality relation \eqref{omega-2}, 
\be
\star_3 \dd \omega_3 = V \dd M - \frac{1}{2} \dd K^3  \implies \omega_3 \simeq \left(\frac{2 }{(1-c)\varkappa^2}m_N - \frac{1}{2}k^3_N\right) \cos \theta_N \dd \phi_N. 
\ee
We have, 
\be \label{omega-simp}
k \simeq \left(\frac{1}{2} k^3_N (1-c) \varkappa^2 + 2 m_N \right) \frac{1}{r_N}\dd \psi_N + \left(\frac{2}{(1-c)\varkappa^2}m_N - \frac{1}{2}k^3_N\right) \cos \theta_N \dd \phi_N. 
\ee
The coefficients in the above equation take values
\bea
 && \frac{1}{2} k^3_N (1-c) \varkappa^2 + 2 m_N   = \frac{(1-c)\sqrt{Q_1 Q_2}\varkappa}{8 \sqrt{2a}} \left((1+a) \sqrt{a-c} + (1-a) \sqrt{a+c}\right) ,\\
&& \frac{2}{(1-c)\varkappa^2}m_N - \frac{1}{2}k^3_N = - \mathrm{i} \frac{\sqrt{Q_1 Q_2}}{8 \sqrt{2a}\varkappa} \sqrt{1-a^2}(\sqrt{a+c} - \sqrt{a-c}).
\eea
From these equations, we note that the coefficient of $\dd \phi_N$ in \eqref{omega-simp} is purely imaginary, while the coefficient of $\dd \psi_N$ is  real. Substituting \eqref{limit-N-1}--\eqref{limit-N-2} and \eqref{omega-simp} into \eqref{6d-string-final} yields the metric in the vicinity of the north circle. It is more convenient to express the metric after performing the following coordinate transformation,
\bea
\sigma_N &=& \frac{\sqrt{Q_1}}{\sqrt{Q_2}}\frac{\sqrt{1-a^2}}{2\sqrt{a}\sqrt{1-c}}(\sqrt{a+c }- \sqrt{a-c})(t + z),  \\
\tau_N &=& \frac{1}{\sqrt{Q_1 Q_2}}\frac{8\sqrt{2a}\varkappa}{\sqrt{1-a^2}(\sqrt{a+ c } - \sqrt{a-c})}  t, \\
\chi_N &=& \sqrt{2} \,   \sqrt{1-c} \, \varkappa \, \psi_N  - \frac{\sqrt{Q_1}}{\sqrt{Q_2}} \frac{(1-a)\sqrt{a+c}+  (1+a)\sqrt{a-c}}{2\sqrt{a}{\sqrt{1-c}}}(t+z).
\eea
In the new coordinates, the metric near the north circle takes the `universal' form, 
\bea
\dd s^2 \simeq &&  \dd \rho_N^2 + \dd \sigma_N^2 + \dd \chi_N^2    + \rho_N^2 (\dd \theta_N^2 + \sin^2 \theta_N \dd \phi_N^2)
- 2 \mspace{2mu} \rho_N \mspace{2mu} \dd \tau_N \dd \sigma_N \nonumber \\
& & 
+ 2 \mspace{2mu} \mathrm{i} \mspace{2mu} \rho_N \mspace{2mu} \cos \theta_N \mspace{2mu} \dd \sigma_N \dd \phi_N.
\eea
The dilaton becomes, 
\be \label{dilaton-north}
e^{2\Phi}  \simeq \frac{4 \sqrt{2}  \varkappa }{\sqrt{1-c}}\frac{g^2  \rho_N}{Q_2},
\ee
and the B-field becomes (upto constant terms which we have dropped),
\be
B \simeq - \rho_N \, \dd \tau_N \wedge \dd \sigma_N - \mathrm{i} \, \rho_N \cos \theta_N \, \dd \sigma_N \wedge \dd \phi_N.
\ee
The metric is singular at the north circle $\rho_N = 0$. The curvature and other field strengths are small only for $\rho_N \gg 1$.

The coordinates $\chi_N, \tau_N, \sigma_N, \phi_N$ have periodic identifications induced from $\psi_N \to \psi_N + 2 \pi, \,  t \to t + \mathrm{i} \beta, \, z \to z + 2 \pi R_z, \phi_N \to \phi_N + 2 \pi$. We will be  specifically interested in the coordinate volume of the two-torus spanned by $\sigma_N$ and  $\chi_N$, which is given as 
\be \label{volume-torus-north}
 4 \pi^2 \frac{\sqrt{Q_1}}{\sqrt{Q_2}}\frac{\sqrt{1-a^2}}{\sqrt{2a}}(\sqrt{a+c }- \sqrt{a-c}) \varkappa R_z.
\ee

\subsubsection{South circle}
This analysis parallels the north circle analysis above with some important differences. To zoom in near the south circle, we  take 
\be
r_S \ll Q_1, Q_2, \varkappa^2. \label{inequalities-S}
\ee
In this limit, the harmonic functions become,
\begin{align}\label{limit-S-1}
&L_1 \simeq  \frac{(1+c)Q_1}{8r_S},&
&L_2 \simeq \frac{(1+c)Q_2}{8r_S}& 
&L_3 = 1, &
&V \simeq \frac{2}{(1+c)\varkappa^2} , \\
&K^1 = 0, & 
&K^2 = 0, & 
&K^3 \simeq \frac{k^3_S}{r_S}, & 
&M \simeq \frac{m_S}{r_S}.  \label{limit-S-2}
\end{align}
Since the function $V$ is now a constant, the four-dimensional base space becomes $\mathbb{R}^3 \times S^1$. The base metric is
\be \label{base-south}
\dd s^2_4 \simeq 2(1+c) \varkappa^2 \dd \psi_S^2 + \frac{2}{(1+c)\varkappa^2} \left(d r_S^2 + r_S^2 \dd \theta_S^2 + r_S^2  \sin^2 \theta_S \dd \phi_S^2\right),
\ee
where $(r_S,\theta_S,\phi_S)$ are spherical polar coordinates centered at $\vec x = \vec x_S$. In writing \eqref{base-south}, we have rescaled the fiber coordinate by a factor of 1/2 and called it $\psi_S$. It has periodicity $2\pi$. Next, we define 
\be
\rho_S = \frac{\sqrt{2}}{\varkappa \sqrt{1+c}} r_S,
\ee
so that the base metric is
\be \label{base-simp-S}
\dd s^2_4 \simeq 2(1+c)\varkappa^2 \dd \psi_S^2 + \dd \rho_S^2 + \rho_S^2 ( \dd \theta_S^2 + \sin^2 \theta_S \dd \phi_S^2).
\ee

Next, following the same procedure as before,  we can compute the one-form $k$ near the south circle. We find,  
\be \label{omega-simp-S}
k \simeq \left(\frac{1}{2} k^3_S (1+c) \varkappa^2 + 2 m_S \right) \frac{1}{r_S}\dd \psi_S + \left(\frac{2}{(1+c)\varkappa^2}m_S - \frac{1}{2}k^3_S\right) \cos \theta_S \dd \phi_S. 
\ee
The coefficients in the above equation take values
\bea
\frac{1}{2} k^3_S (1+c) \varkappa^2 + 2 m_S  &=& \frac{(1+c)\sqrt{Q_1 Q_2}\varkappa}{8 \sqrt{2a}} \left((1+a) \sqrt{a+c}+ (1-a) \sqrt{a-c} \right) ,\\
\frac{2}{(1+c)\varkappa^2}m_S - \frac{1}{2}k^3_S &=& \mathrm{i} \frac{\sqrt{Q_1 Q_2}}{8 \sqrt{2a}\varkappa} \sqrt{1-a^2}(\sqrt{a+c} - \sqrt{a-c}).
\eea
The coefficient of $\dd \phi_S$ in \eqref{omega-simp-S} is purely imaginary (and is the complex conjugate of the corresponding coefficient at the north circle), while the coefficient of $\dd \psi_S$ is real.
 Substituting \eqref{limit-S-1}--\eqref{limit-S-2} and \eqref{omega-simp-S} into \eqref{6d-string-final} yields the metric in the vicinity of the south circle. It is more convenient to express the metric after performing the following coordinate transformation,
\bea
\sigma_S &=& \frac{\sqrt{Q_1}}{\sqrt{Q_2}}\frac{\sqrt{1-a^2}}{2\sqrt{a}\sqrt{1+c}}(\sqrt{a+c }- \sqrt{a-c})(t + z),  \\
\tau_S &=& \frac{1}{\sqrt{Q_1 Q_2}}\frac{8\sqrt{2a}\varkappa}{\sqrt{1-a^2}(\sqrt{a+ c } - \sqrt{a-c})}  t, \\
\chi_S &=& \sqrt{2} \,   \sqrt{1+c} \, \varkappa \, \psi_S  - \frac{\sqrt{Q_1}}{\sqrt{Q_2}} \frac{(1+a)\sqrt{a+c}+  (1-a)\sqrt{a-c}}{2\sqrt{a}{\sqrt{1+c}}}(t+z).
\eea
In the new coordinates, the metric near the south circle takes the `universal' form,
\bea
ds^2 \simeq &&  \dd \rho_S^2 + \dd \sigma_S^2 + \dd \chi_S^2  + \rho_S^2 (\dd \theta_S^2 + \sin^2 \theta_S \dd \phi_S^2)- 2 \mspace{2mu} \rho_S \mspace{2mu} \dd \tau_S \dd \sigma_S
\nonumber \\
& & 
 -  2 \mspace{2mu} \mathrm{i} \mspace{2mu} \rho_S \mspace{2mu} \cos \theta_S \mspace{2mu} \dd \sigma_S \dd \phi_S.
\eea
Note the change in the minus sign in the $\dd \sigma_S \dd \phi_S$ term compared to the north circle term $\dd \sigma_N \dd \phi_N$. Otherwise the metrics are the same. 
The dilaton becomes, 
\be \label{dilaton-south}
e^{2\Phi}  \simeq \frac{4 \sqrt{2}  \varkappa }{\sqrt{1+c}}\frac{g^2  \rho_S}{Q_2},
\ee
and the B-field becomes (upto constant terms which we have dropped),
\be
B \simeq - \rho_S \, \dd \tau_S \wedge \dd \sigma_S + \mathrm{i} \, \rho_S \cos \theta_S \, \dd \sigma_S \wedge \dd \phi_S.
\ee
Once again, note the change in the minus sign in the $\dd \sigma_S \wedge \dd \phi_S$ term compared to the north circle term $\dd \sigma_N \wedge \dd \phi_N$.
The metric is singular at the south circle $\rho_S = 0$. The curvature and other field strengths are small only for $\rho_S \gg 1$. The `universal' form of the metric and the B-field at the south circle are essentially the same as at the north circle.  The minus sign differences can be removed by defining $\sigma_S' = - \sigma_S$ and $\tau_S' = - \tau_S$.

The coordinates $\chi_S, \tau_S, \sigma_S, \phi_S$ have periodic identifications induced from $\psi_S \to \psi_S + 2 \pi, \,  t \to t + \mathrm{i} \beta, \, z \to z + 2 \pi R_z, \phi_S \to \phi_S + 2 \pi$. We will be  specifically interested in the coordinate volume of the two-torus spanned by $\sigma_S$ and  $\chi_S$, which is given as 
\be \label{volume-torus-south}
 4 \pi^2 \frac{\sqrt{Q_1}}{\sqrt{Q_2}}\frac{\sqrt{1-a^2}}{\sqrt{2a}}(\sqrt{a+c }- \sqrt{a-c}) \varkappa R_z.
\ee
Note that this expression is the same as the corresponding expression at the north circle, cf.~\eqref{volume-torus-north}.

\subsection{Adding the two contributions}
\label{sec:limitations}

The parameters $Q_1,Q_2$ and $q$  are related to the quantized charges $n,w,Q$ via the relations \eqref{quantized}. The fourth relation in \eqref{quantized} involving $\varkappa^2$, $J_\psi$, and $q$ hinges on \eqref{J-psi-small-black-ring}.  As noted below \eqref{dipole-alpha-2}, the angular momentum $J_\psi$ of the index saddle and the dipole charge $q$ satisfy \eqref{J-psi-small-black-ring} only when terms of $\mathcal{O}(a^2)$ are neglected.

With these observations in mind, we substitute $c = a \alpha$ in \eqref{volume-torus-north} and \eqref{volume-torus-south} and expand in powers of $a$. We find
\be
 2 \sqrt{2} \pi^2 \frac{\sqrt{Q_1}}{\sqrt{Q_2}}(\sqrt{1+\alpha }- \sqrt{1-\alpha}) \varkappa R_z + \mathcal{O}(a^2), 
\ee
which can be written as
\be \label{area-sigma-chi-A}
A_{\sigma \chi} = 4 \pi^2 \sqrt{\frac{J n}{Qw}}\sqrt{1 - \frac{JQ}{nw}} + \mathcal{O}(a^2).
\ee
There are no $\mathcal{O}(a)$ terms in these expansions. Moreover, all dependence on the moduli $g$ and $R_z$ has canceled at the leading order.  Since the coordinate volumes of the two-tori at the north and south circles are same, we have dropped the sub-scripts $N$ and $S$ and have denoted the coordinate volume as $A_{\sigma \chi}$.

  In the large charge limit, the string coupling is small near the singularities. Thus, we can ignore the string loop corrections near the singularities.  The string tree level action involving the string frame metric $G_{\mu \nu}$, the dilaton $\Phi$, and the anti-symmetric $B_{\mu \nu}$ field has the form, 
\be
\mathcal{S} = \int d^6x \sqrt{-\det G} e^{-2\Phi} \mathcal{L} \label{effective-action}
\ee
where $\mathcal{L}$ is a function of the metric, Riemann tensor, the field strength of the B-field, and covariant derivatives of the dilaton $\Phi$, but not of the dilaton itself. Since the graviational index vanishes at the two-derivative level, cf.~\eqref{S-J}, the net contribution to the index due to higher-derivative corrections can be computed from the on-shell action near the singularities \cite{Chowdhury:2024ngg}.

We saw that near the singularities the solution takes a universal form. The universal form is independent of all coordinates except the radial and polar coordinates. We expect higher derivative corrections to the solution will continue to preserve these symmetries near the circles.   Near the north circle, the dilaton takes the form \eqref{dilaton-north}. As a result,  the dilaton term $e^{-2\Phi}$ in \eqref{effective-action} gives rise to a multiplicative factor
\be
\frac{1}{4}\sqrt{1-c} \frac{ Q_2}{\sqrt{2} \kappa g^2} = \frac{1}{4}\sqrt{1-a \alpha} \left( w  \sqrt{\frac{Q}{J}} \right) + \mathcal{O}(a^2),
\ee
as an overall normalization in the higher-derivative corrected effective action from the north circle. From the coordinate area   of the $ \sigma, \chi$ torus we get another multiplicative factor \eqref{area-sigma-chi-A}. 
Now, as per the reasoning of \cite{Chowdhury:2024ngg,Dabholkar:2006za}, the contribution to the logarithm of the index from the north circle is therefore, 
\be
\frac{1}{4}K \sqrt{1-a \alpha} \left( w \,  \sqrt{\frac{Q}{J}} \right)  A_{\sigma \chi} + \mathcal{O}(a^2),
\ee
and from the south circle is, 
\be
\frac{1}{4}K \sqrt{1+a \alpha} \left( w \,  \sqrt{\frac{Q}{J}} \right)  A_{\sigma \chi} + \mathcal{O}(a^2),
\ee
where $K$ is a constant that captures the effect on the higher-derivative corrections. It is the same constant that appears in the two contributions. This is because the geometry near the north and south circles are identical. Adding the two contributions we get
\be
\frac{1}{4} K\left(\sqrt{1-a \alpha } + \sqrt{1+a \alpha} \right) \left( w \sqrt{\frac{Q}{J}} \right) A_{\sigma \chi} + \mathcal{O}(a^2). 
\ee
Now, the key point is that in the sum $ \left(\sqrt{1-a \alpha } + \sqrt{1+a \alpha} \right) $ the term linear in $a$ cancels in an expansion near $a=0$. As a result, the corrected gravitational index to order $\mathcal{O}(a)$ is 
\be
 \left( \frac{1}{2} K  w \sqrt{\frac{Q}{J}} \right)  \times \left( 4 \pi^2 \sqrt{\frac{J n}{Qw}}\sqrt{1 - \frac{JQ}{nw}} \right) + \mathcal{O}(a^2) 
= 2 \pi^2 K \sqrt{nw - J Q}+ \mathcal{O}(a^2).
\ee

Several contributions combine to produce a precise cancellation at $\mathcal{O}(a)$. A possible justification for the small $a$ expansion proceeds as follows. The two centers $\vec x_N$ and $\vec x_S$ in the Bena--Warner description are separated by the distance
\be
|\vec x_N - \vec x_S| = c \varkappa^2 = a \alpha \varkappa^2 .
\ee
For $a \sim c \sim 1$, this suggests that when $\varkappa^2 \gg 1$, the index-saddle black ring corresponds to a relatively ``fat'' black ring, with the distance between the centers scaling with $\varkappa$ in the same way as the size of the $S^1$ of the ring. By contrast, when $a$ is  sufficiently small, an additional scale emerges that controls the shape of the ring. It is only in this ``thin'' black ring regime that the scaling analysis correctly reproduces the microscopic index.

We conclude this section by  highlighting the regime of validity of our scaling analysis. The analysis presented applies only within a restricted region of parameter space, whose full significance is not yet fully understood within our construction. In particular, several distinct contributions combine to produce a precise cancellation at order 
$\mathcal{O}(a)$. While the deeper origin of this cancellation remains unclear, it is nevertheless remarkable that, within this limited regime, the scaling analysis successfully reproduces the expected microscopic behavior of the index.  We also recall that, despite significant effort, the index saddles for small black holes in four dimensions remain poorly understood \cite{Chowdhury:2024ngg, Chen:2024gmc, Hegde:2024bmb}. Since the small black ring studied in this paper is closely tied to the four-dimensional small black hole, it is not surprising that certain gaps remain in the analysis and understanding. A more complete understanding of this structure, and of the role played by higher-derivative corrections beyond the present approximation, would be an important direction for future work.

\section{Conclusions}
\label{sec:conclusions}

In this work, we constructed the gravitational index saddle for the supersymmetric F1–P black ring by taking a  supersymmetric limit of a recently obtained~\cite{our-paper} non-extremal doubly spinning F1–P black ring. The resulting index saddle can be written naturally as a three-center Bena–Warner configuration and carries all four charges $(n,w,J,Q)$ of the small black ring and an additional parameter. This extends the construction of \cite{Bandyopadhyay:2025jbc}, where the dipole charge was not an independent parameter.

Although the saddle solution possesses a finite-area event horizon, the gravitational index vanishes at the two-derivative level. This mirrors the behavior of small black holes \cite{Chowdhury:2024ngg, Chen:2024gmc} and underscores the necessity of higher-derivative corrections \cite{Hegde:2024bmb, Adhikari:2025eok}. Indeed, the index saddle is not completely smooth: it is singular on two circles of the horizon where the string coupling vanishes, indicating the breakdown of the two-derivative supergravity approximation. By analyzing the geometry near these circles and exploiting the symmetries of the string-frame equations of motion, we attempted a scaling analysis following \cite{Chowdhury:2024ngg, Dabholkar:2006za}.

Within a restricted region of parameter space, which as we argued corresponds to sufficiently thin black rings, we found that the corrected index reproduces the expected microscopic behavior, up to an overall numerical constant that cannot be fixed by scaling arguments alone. The appearance of this restricted regime, as well as the role played by the additional parameter of the index saddle, point to subtleties that remain to be fully understood. Similar subtleties remain to be fully understood in relation to the corresponding four-dimensional small black hole index saddles.

We view our results as further evidence that the index saddle framework provides a gravitational description of protected microscopic quantities, even in situations where the classical black hole solution is singular. It would be interesting to extend this analysis to include a more systematic treatment of higher-derivative corrections.  Finally, connecting our results to recent constructions of bubbling and multi-centered index saddles \cite{Boruch:2025biv, Cassani:2025iix, Boruch:2025sie}, may reveal deeper structural insights into semiclassical index computations.

\bigskip
\noindent \textbf{Acknowledgments:} The work of A.V. was partly supported by SERB Core Research Grant CRG/2023/000545. The work of G.S.P. was supported by the National Natural Science Foundation of China (NSFC) under Grant No.~12247103. A.V. dedicates this paper to the memory of his father, late Shri Ramesh Kumar Virmani, who passed away after a courageous battle with cancer.

\appendix
\section{Alternative set of ring coordinates}
\label{sec:appendix}
The following coordinate transformation on flat base space \eqref{r1r2-base}  yields an alternative set of ring coordinates $(\bar x, \bar y)$: 
\begin{align} \label{r1r2}
r_1^2 &=\frac{R^2 \left(1-\bar x^2\right) \left(1 - \bar b^2 
   \bar  y^2\right)}{(1+ \bar b^2 ) (\bar  x-\bar  y)^2}, &
r_2^2 &=\frac{R^2 \left(\bar  y^2-1\right) \left(1 - \bar b^2 
   \bar  x^2\right)}{(1 + \bar b^2 ) (\bar x-\bar y)^2}.
\end{align}
The $(\bar x, \bar y)$ coordinates are used in \cite{Bandyopadhyay:2025jbc} and are related to the $(x,y)$ coordinates used in Section \ref{sec:index-saddle} as,
\begin{align}
c &= \frac{2 \bar b }{1+ \bar b^2},&
x &= \frac{\bar x - \bar b }{1 - \bar b  \bar x }, &
y &= \frac{\bar y - \bar b }{1 - \bar b  \bar y }. \label{parameter-change}
\end{align}
In \cite{Bandyopadhyay:2025jbc}, the charge parameters $Q_1$ and $Q_2$ were normalized differently. Specifically, 
\begin{align} \label{canonical-charges}
{\textbf Q}_1 &= \frac{\pi }{4 G_5 }  \frac{Q^\mathrm{there}_1}{1 - \bar b^2}, & {\textbf Q}_2 &= \frac{\pi }{4 G_5 }  \frac{Q^\mathrm{there}_2}{1 - \bar b^2}. 
\end{align}
Therefore, the relation between the charge parameters $Q_1$ and $Q_2$ used in this paper and those used in \cite{Bandyopadhyay:2025jbc} is
\be
Q^\mathrm{there}_{1,2} = (1-\bar b^2) Q_{1,2}.
\ee
Setting $a=c$ in the harmonic functions of Section \ref{sec:BW} and replacing $c$ by $\bar b$ via \eqref{parameter-change}, we find that these functions match  perfectly with the corresponding harmonic functions presented in Section 6 of \cite{Bandyopadhyay:2025jbc}. 

We now present two observations that clarify the
 $b \to -1$ limit mentioned at the beginning of Section \ref{sec:index-saddle}. Admittedly, the reader may find these comments cryptic; they are not needed for any of the arguments presented in the paper. These observations helped us develop intuition for taking the BPS limit, which involved taking $b\to-1$. 
\begin{enumerate}
\item Section 6.5 of reference \cite{Chen:2012kd} discusses how to take the infinite radius limit of the doubly spinning dipole black ring. In this limit, one obtains a charged rotating black string in standard Boyer-Lindquist coordinates. One of the transformations given in \cite[eq.~(6.4)]{Chen:2012kd} is
\be
b = \frac{m - \sqrt{m^2 - a_K^2}}{m + \sqrt{m^2 -a _K^2}}, \label{para-b-aK}
\ee
where $m$ and $a_K$ are the  mass and rotation parameters of the Kerr black hole, respectively. 

For the sake of the argument, consider constructing the index saddle using the black hole obtained by dimensional reduction of this black string. To construct index saddles in related situations, such as  \cite{Chowdhury:2024ngg}, we consider the analytic continuation $b_K = \mathrm{i} a_K$ and take  the BPS limit $m \to 0$ while keeping $b_K$ fixed. In this limit, it follows immediately from \eqref{para-b-aK} that  the parameter $b$ approaches $-1$. This observation suggests that, in constructing index saddles from the non-extremal black ring, one should consider the limit $b \to -1$.

    \item The non-extremal charged solution in~\cite{Bandyopadhyay:2025jbc} is constructed by applying boosts and T-duality to the Pomeransky--Sen'kov black ring \cite{Pomeransky:2006bd}. As discussed in Section 4.4.2 of our previous paper \cite{our-paper}, the transformation that relates the \emph{non-extremal} $a=c$ solution of \cite{our-paper} to that of  \cite{Bandyopadhyay:2025jbc} is,
\begin{align}
b&=\frac{\bar \nu(1-\bar \mu^2)}{\bar \mu(1-\bar \nu^2)}\,,& c&=\frac{\bar \mu-\bar \nu}{1-\bar \mu \bar \nu}\,, \label{para1} \\[2mm]
 x&=\frac{\bar{x}+\bar \nu}{1+ \bar \nu \bar {x}}\,,& y&=\frac{\bar{y}+\bar \nu}{1+\bar \nu \bar{y}}\,, \label{coordinate-x-barx-y-bary}
\end{align}
where the coordinates $\bar{x}$ and $\bar{y}$ are identified with the coordinates $x$ and $y$ of \cite{Bandyopadhyay:2025jbc}, respectively, and the parameters  $\bar \mu$ and $\bar \nu$ are related to parameters $\nu$ and $\eta$ used in \cite{Bandyopadhyay:2025jbc} via,
\bea
\nu &=& \bar \mu + \bar \nu, \label{para2} \\
\eta &=&  \bar \mu \, \bar \nu. \label{para3}
\eea
The BPS limit discussed in \cite{Bandyopadhyay:2025jbc} involved two steps: (i) continuing $\eta$ to negative values, $\eta = - \bar b^2$, with $0 \le \bar b \le 1$, and (ii) taking the parameter $\nu \to 0$. 
The necessity of taking 
$\nu \to 0$ is evident from the ADM mass expressions given in \cite{Pomeransky:2006bd, Bandyopadhyay:2025jbc}. Using \eqref{para1}–\eqref{para3}, it is straightforward to see that this BPS limit corresponds to taking 
$b \to -1$. This independently supports the conclusion that in constructing index saddles from the non-extremal black ring, one should consider the limit 
$b \to -1$. We also note that in the $\nu \to 0$ limit, transformations \eqref{coordinate-x-barx-y-bary} are the same as \eqref{parameter-change} with $\bar \mu = - \bar \nu = \bar b.$.
\end{enumerate}

\bibliography{dipole-black-ring-2}
\bibliographystyle{JHEP}

\end{document}